\documentclass[twocolumn]{autart}

\usepackage{graphicx}
\usepackage{amsfonts}
\usepackage{amssymb}
\usepackage{enumerate}
\usepackage{color}
\usepackage[tbtags,fleqn]{amsmath}
\usepackage{indentfirst}
\usepackage{subcaption}
\allowdisplaybreaks[4]
\begin{document}

\begin{frontmatter}

\title{Practical Output Consensus of Nonlinear Heterogeneous Multi-Agent Systems with Limited Data Rate \vspace{-18pt}}


\thanks[footnoteinfo]{This paper was not presented at any IFAC meeting. Corresponding author Lihua Xie.}

\author[]{Maopeng Ran}\ead{mpran@ntu.edu.sg},
\author[]{Lihua Xie}\ead{ELHXIE@ntu.edu.sg}

\address{School of Electrical and Electronic Engineering, Nanyang Technological University, Nanyang Avenue, Singapore 639798, Singapore \vspace{-18pt}}


\begin{abstract}

This paper investigates the consensus problem for nonlinear heterogeneous multi-agent systems with limited communication data rate. Each agent is modeled by a higher-order strict-feedback continuous-time system with unknown nonlinearities and external disturbance, and only the first state variable being measurable. Extended state observers (ESOs) are used to estimate the unmeasurable agent states and the unknown nonlinear dynamics. An ESO-based distributed output feedback protocol with dynamic encoding and decoding is then presented. It is shown that, for a connected undirected network, the proposed protocol guarantees practical output consensus, in which the steady-state consensus error can be made arbitrarily small. The ESO-based protocol also shapes the transient consensus performance, as it is capable of recovering the consensus performance of a linear counterpart with fully measurable states. Furthermore, we prove that for higher-order uncertain nonlinear multi-agent systems, consensus can be achieved with merely one bit information exchange between each pair of adjacent agents at each time step. Finally, simulations on third-order pendulum systems are given, which verify the theoretical results.

\vspace{-12pt}

\end{abstract}

\end{frontmatter}

\section{Introduction}

Multi-agent systems have attracted great attention in recent years owing to their potential applications in a wide variety of tasks, such as exploration, surveillance, and cooperative manipulation. Consensus, which aims at all agents autonomously achieving an agreement of common interest by using local information available at the node and received from neighboring agents, is one of the most fundamental problems in multi-agent systems.

{In most cases, the agents communicate with their neighbours through digital networks. For a multi-agent system with a digital communication network, the information to be exchanged between each pair of adjacent agents will be first quantized by the sender and then sent out to its neighbours. When the neighbours receive the quantized information through the digital communication channels, they use a decoding algorithm to reconstruct the information. That is to say, only quantized information can be exchanged between agents. Quantized consensus for first-order multi-agent systems was investigated in Kashyap, Basar, and Srikant (2007), Carli, Bullo, and Zampieri (2010), Li, Fu, Xie, and Zhang (2011), and Mu and Liu (2015). Specifically, Li et al. (2011) studied the average consensus problem for undirected networks of first-order discrete-time integrator systems under finite bit-rate communication. A static uniform quantizer with an exponentially decaying scaling function was proposed, and it was shown in Li et al. (2011) that no matter how many agents there are, the consensus can be achieved with as few as one bit information exchange between each pair of adjacent agents at each time step.

For higher-order multi-agent systems, the limited data rate problem becomes more challenging. Li and Xie (2012) developed a quantized-observer based encoding-decoding scheme for second-order integrator systems with partially measurable states. It was shown in Li and Xie (2012) that for a connected network, two bits information exchange between each pair of adjacent agents at each time step suffices to guarantee the consensus of the agents. For multi-agent systems with $r$th-order integrator dynamics, Qiu, Xie, and Hong (2017a)  showed that $r$ bits of information exchange between each pair of adjacent agents at each time step are necessary to guarantee consensus. Qiu, Xie, and Hong (2017b) further investigated the data rate problem for $2r$th-order multi-agent systems with oscillator dynamics, and the sufficient number of communication bits to guarantee consensus was proved to be an integer between $r$ and $2r$, depending on the location of the poles of the agent state matrix. You and Xie (2011) and Meng, Li, and Zhang (2017) considered the limited data rate problem for general linear multi-agent systems with fully and partially measurable states, respectively. However, the minimum data rate required in You and Xie (2011) and Meng et al. (2017) remains an open problem.

The works mentioned above mainly focused on linear multi-agent systems. In a recent work, Dong (2019) first visited the limited data rate problem for a class of nonlinear multi-agent systems with fully measurable states. In the case that there exist a known model of the agent nonlinearity and a known upper bound of the modeling error, a multi-loop control structure was proposed in Dong (2019) to guarantee consensus. In Ran and Xie (2019), the limited data rate problem was considered for a class of simple nonlinear multi-agent systems without external disturbances.

Based on the above observations, there are several important and interesting problems that remain unsolved. Firstly, it still lacks a practical solution to handle the limited data rate problem for general multi-agent systems with nonlinearities, uncertainties, and external disturbances. It should be pointed out that the nonlinearities, uncertainties, and external disturbances exist in almost all practical control systems (Khalil, 2002; Isidori, 1989). Secondly, whether the data rates required in Li and Xie (2012) and Qiu et al. (2017a; 2017b) are conservative, and how many bits are sufficient for general higher-order uncertain nonlinear multi-agent systems with partially measurable states? Thirdly, for the  limited data rate problem, can a comparable transient consensus performance for a multi-agent system  with nonlinearities, uncertainties, and external disturbances, to its simple linear counterpart be achieved?

In this paper, we investigate the limited data rate problem for nonlinear heterogeneous multi-agent systems with partially measurable states. Each agent is described by a strict-feedback system with unknown nonlinear dynamics and external disturbance. Extended state observer (ESO) (Han, 2009; Khalil, 2017)  is the main tool used in this paper. The unknown dynamics is first regarded as an extended state of the agent, and then the agent state and the defined extended state are estimated simultaneously by the ESO. An ESO-based protocol with dynamic encoding and decoding is then presented. Comparing with the existing literature, the main contributions of this paper are threefold:
\begin{enumerate}[i)]
  \item We provide a more practical solution to the limited data rate problem for multi-agent systems. The proposed protocol is output-feedback-based and  is capable of handling nonlinearities, uncertainties, and external disturbances.  In practical applications, the agent states may be not fully available for feedback; and its dynamics is very likely to be perturbed by uncertainties and external disturbances (Khalil, 2002; Isidori, 1989).
  \item We show that, with the application of the proposed ESO-based protocol, merely \emph{one bit} information exchange between each pair of adjacent agents at each time step suffices to guarantee the output consensus of general higher-order uncertain nonlinear multi-agent systems with partially measurable states. This indicates that the data rates required in Li and Xie (2012) and Qiu et al. (2017a; 2017b) are still conservative, and from a theoretical viewpoint, the approach developed in this paper achieves the lowest data rate required at each time step via output feedback.
  \item The proposed ESO-based protocol is capable of shaping the transient consensus performance of the uncertain nonlinear multi-agent systems with limited data rate.   More specifically, the agent trajectories of an uncertain nonlinear multi-agent system under the ESO-based output feedback protocol can be made arbitrarily close to the trajectories of its linear counterpart with fully measurable states.
\end{enumerate}

\section{Problem Formulation}

\subsection{Notation}

Let $\mathbb{R}$ and  $\mathbb{R}^{n}$ denote the sets of real numbers and $n$-dimension real vectors, respectively. $I_n$ and $\textbf{0}_n$ represent the identity matrix and the vector of zeros of dimension $n$, respectively. Let $\textbf{1}_N$ be the $N$-dimensional column vector with all components being 1, and define $J_N=(1/N)\textbf{1}_N\textbf{1}^{\textrm{T}}_N$. $\textrm{diag}\{a_1,\ldots, a_n\}$ is the diagonal matrix with the $i$-th diagonal component equal to $a_i$. For a given vector or matrix $A$, $A^{\rm{T}}$, $\|A\|$, and $\|A\|_{\infty}$ represent its transpose, Euclidean norm, and $\infty$-norm, respectively. For a given positive number $a$, $\lfloor a  \rfloor$ and $\lceil a \rceil$ represent the largest integer not greater than $a$ and the smallest integer not less than $a$, respectively.  Big $O$-notation in terms of $\nu$ is denoted as $O(\nu)$ and it is assumed to hold for a sufficiently small positive $\nu$. Let $\textrm{sat}(\cdot)$ represent the unity saturation function defined by $\textrm{sat}(\nu)=\textrm{sign}(\nu)\cdot\min\{1, |\nu|\}$. Throughout this paper, for clear presentation, the time variable $t$ of a signal will be omitted except when the dependence of the signal on $t$ is crucial.

\subsection{Communication Graph}

The communications between agents are modeled as a graph $\mathcal{G}=\{\mathcal{V},\mathcal{E},\mathcal{A}\}$, where $\mathcal{V}=\{1, 2, \ldots, N\}$ is the index set of $N$ agents with $i$ representing the $i$th agent, $\mathcal{E}\subset\mathcal{V}\times\mathcal{V}$ is the edge set of paired agents, and $\mathcal{A}=[a_{ij}]\in\mathbb{R}^{N\times N}$ is the weighted adjacency matrix with $a_{ij}=1$ or $0$ indicating whether or not there is a communication channel from agent $j$ to $i$. If $a_{ij}=a_{ji}$ for any pair of neighbouring agents, the associated communication graph is called an undirected graph. The neighborhood of the $i$th agent is represented by $\mathcal{N}_i=\{j\in\mathcal{V}|(i,j)\in\mathcal{E}\}$. The in-degree of agent $i$ is denoted as $\textrm{deg}_i=\sum_{j=1}^{N}a_{ij}$, and $d^*=\max_{1\leq i\leq N}\textrm{deg}_i$ is the degree of $\mathcal{G}$. The Laplacian matrix of $\mathcal{G}$ is defined as $\mathcal{L}=\mathcal{D}-\mathcal{A}$, where $\mathcal{D}=\textrm{diag}\{\textrm{deg}_1,\ldots, \textrm{deg}_N\}$. A sequence of edges $(i_1,i_2)$, $(i_2,i_3)$, $\ldots$, $(i_{k-1},i_k)$ is called a path from agent $i_1$  to agent $i_k$. The graph $\mathcal{G}$ is a connected graph if for any two agents $i, j\in\mathcal{V}$, there exists a path from $i$ to $j$. For a connected graph, its Laplacian matrix $\mathcal{L}$ is a symmetric positive semidefinite matrix and its eigenvalues in an ascending order are denoted by $0=\lambda_1(\mathcal{L})< \lambda_2(\mathcal{L})\leq\cdots \leq \lambda_N(\mathcal{L})$ (Fiedler, 1973).
\vspace{8pt}

\subsection{Problem Statement}
We consider a multi-agent system consisting of $N$ nonlinear agents described by
\begin{equation}\label{eq1}
 \left\{
  \begin{aligned}
          \dot{x}_{im}=& f_{im}(x_{i1},\ldots,x_{im})+x_{i,m+1}, 1\leq m\leq r-1,\\
          \dot{x}_{ir}=& f_{ir}(x_i, z_i,\omega_i)+u_i, \\
          \dot{z}_i  =& f_{i0}(x_i,z_i,\omega_{i}), \\
                y_i=& x_{i1}, ~i=1, \ldots, N,
        \end{aligned} \right.
\end{equation}
where $x_i=[x_{i1}, x_{i2}, \ldots, x_{ir}]^{\rm{T}}\in\mathbb{R}^r$ is the agent state, $z_i\in\mathbb{R}^{n_i-r}$ is the state of the zero dynamics, $n_i$ is the dimension of each agent, $r$ is the relative degree,
$y_i\in\mathbb{R}$ is the measured output, $u_i\in\mathbb{R}$ is the control input, $\omega_{i}\in\mathbb{R}^{n_{\omega_{i}}}$ is the external disturbance, $f_{i0}\in C^1\left(\mathbb{R}^r \times \mathbb{R}^{n_i-r}\times\mathbb{R}^{n_{\omega_{i}}},\mathbb{R}^{n_i-r}\right)$,  $f_{im}\in C^{r+1-m}\left(\mathbb{R}^{m}, \mathbb{R}\right)$, $1\leq m\leq r-1$, and $f_{ir}\in C^{1}\left(\mathbb{R}^{r}\right.$ $\left. \times \mathbb{R}^{n_i-r}\times \mathbb{R}^{n_{\omega_{i}}}, \mathbb{R}\right)$ are locally Lipschitz functions.
\vspace{6pt}
\\
\textbf{Remark 1.} System (\ref{eq1}) covers the plants considered in the existing data rate literature as special cases. Specifically, the first-order integrator considered in Li et al. (2011) and Mu and Liu (2015), the second-order integrator considered in Li and Xie (2012), and the higher-order dynamics considered in You and Xie (2011), Ment et al. (2017), Qiu et al. (2017a; 2017b), Dong (2019), and Ran and Xie (2019), directly or under a well-defined state transformation, fall into the form of system (\ref{eq1}). What is more, since the functions $f_{i0}(\cdot), f_{i1}(\cdot), \ldots, f_{ir}(\cdot)$ can be different for different agents, the multi-agent system described by (\ref{eq1}) is heterogeneous.
\vspace{6pt}

For system (\ref{eq1}), we have the following assumptions:
\vspace{6pt}
\\
\textbf{Assumption A1.} The  external disturbance $\omega_i$ and its derivative $\dot{\omega}_i$ are bounded.
\vspace{6pt}
\\
\textbf{Assumption A2.} The zero dynamics $\dot{z}_i=f_{i0}(x_i,z_i,\omega_{i})$, with input $(x_i,\omega_{i})$, is bounded-input-bounded-state (BIBS) stable.
\vspace{6pt}
\\
\textbf{Assumption A3.} The communication graph $\mathcal{G}$ among the $N$ agents is undirected and connected.
\vspace{6pt}

In this paper, we will investigate the consensus of the multi-agent system (\ref{eq1}) with a digital network. Signals to be transmitted between each pair of adjacent agents are first quantized and encoded at the transmitters, and then decoded at the receivers. The finite-level uniform quantizer employed in this paper is given by
\begin{equation}\label{eq20}
 q(\nu)=\left\{
\begin{aligned}
&0, && -1/2<\nu<1/2, \\
&i, && \frac{2i-1}{2} \leq \nu <\frac{2i+1}{2}, \\
&&&  i=1, 2, \ldots, K-1, \\
&K, && \nu \geq \frac{2K-1}{2}, \\
&-q(-\nu), && \nu \leq -1/2,
\end{aligned} \right.
\end{equation}
where $K\in\{1,2,\ldots,\}$, and $2K+1$ is the quantization level. Similar to Li et al. (2011), it is assumed that the agent sends out no signal if the output of the quantizer is 0. Therefore, for a $(2K+1)$-level quantizer $q(\cdot)$, it suffices to use $\lceil\log_2(2K)\rceil$ bits to represent its output. In particular, letting $K=1$ yields a one-bit quantizer.

In this paper, we aim to solve the output consensus problem for the nonlinear heterogeneous multi-agent system (\ref{eq1}) with limited data rate. For simplicity and to illustrate our design scheme, we first consider the simple case that each agent has full knowledge of its dynamics (i.e., $x_i$, $z_i$, $\omega_i$, and $f_{im}(\cdot)$, $0\leq m\leq r$) in Section 3. Then, in Section 4,  we move forward to the general case that $\omega_i$ and $f_{im}(\cdot)$, $0\leq m\leq r$, are uncertain, and each agent only has access to its first state variable.

\section{Consensus Under Full Information}

In this section, based on the full information of the agent dynamics, new state variables $\varrho_i=[\varrho_{i1},\ldots,\varrho_{ir}]^{\rm{T}}\in\mathbb{R}^r$, $1\leq i\leq N$, are first introduced for subsequent consensus protocol design. Let $\varrho_{i1}=y_i=x_{i1}\triangleq \phi_{i1}(x_{i1})$, $\varrho_{im}=\dot{\varrho}_{i, m-1}=\sum_{j=1}^{m-2}(x_{i,j+1}+f_{ij}(x_{i1},\ldots,x_{ij}))\frac{\partial \phi_{i, m-1}}{\partial x_{ij}}+x_{im}+f_{i,m-1}(x_{i1},\ldots,x_{i, m-1}) \triangleq \phi_{im}(x_{i1},\ldots,x_{im})$, $2\leq m\leq r$, and $\dot{\varrho}_{ir}=  f_{ir}(x_i,z_i,\omega_{i})+u_i
  +\sum_{j=1}^{r-1}(x_{i,j+1}+f_{ij}(x_{i1},\ldots,x_{ij}))\frac{\partial \phi_{ir}}{\partial x_{ij}}
\triangleq  F_i(\varrho_i,z_i,\omega_i)+u_i$, where $F_i\in C^1(\mathbb{R}^{r}\times \mathbb{R}^{n_i-r}\times \mathbb{R}^{n_{\omega_i}}, \mathbb{R})$. As a result, the original system (\ref{eq1}) is transformed into the following canonical form (Isidori, 1989):
\begin{equation}\label{eq27}
 \left\{
  \begin{aligned}
          \dot{\varrho}_i=& A\varrho_i+B[F_i(\varrho_i, z_i, \omega_i)+u_i], \\
                    \dot{z}_i  =& Z_i(\varrho_i,z_i,\omega_i), \\
                y_i=& \varrho_{i1}, ~i=1, \ldots, N,
        \end{aligned} \right.
\end{equation}
where $Z_i\in C^1(\mathbb{R}^{r}\times \mathbb{R}^{n_i-r}\times \mathbb{R}^{n_{\omega_i}}, \mathbb{R}^{n_i-r})$ represents the zero dynamics with respect to the new state variable $\varrho_i$, and matrices  $A=
\left[
  \begin{array}{cc}
    \textbf{0} _{r-1} & I_{r-1} \\
    0   & \textbf{0}^{\textrm{T}}_{r-1}
    \end{array}\right]$ and $B=\left[
             \begin{array}{c}
               \textbf{0}_{r-1} \\
               1 \\
             \end{array}
           \right]$.
Since $\varrho_{i1}=x_{i1}$, the output consensus of system (\ref{eq27}) is equivalent to the output consensus of system (\ref{eq1}).
\vspace{6pt}
\\
\textbf{Lemma 1.} \emph{Let
\begin{equation}\label{eq2}
s_i=k_1\varrho_{i1}+k_2\varrho_{i2}+\cdots+k_{r-1}\varrho_{i, r-1}+\varrho_{ir},
\end{equation}
where $k_1, k_2,\ldots, k_{r-1}$  are chosen such that the polynomial $k_1+k_2\lambda+\cdots+k_{r-1}\lambda^{r-2}+\lambda^{r-1}$ is Hurwitz.
\begin{itemize}
  \item If $s_i$ is bounded and $\lim_{t\rightarrow \infty}s_i=\Lambda$ for some constant $\Lambda$, then the multi-agent system (\ref{eq1}) achieves the following output consensus:
\begin{equation}\label{eq29}
\lim_{t\rightarrow \infty}(y_{i}-y_{j})=0, ~ 1\leq i\neq j\leq N.
\end{equation}
Furthermore, if the convergence of $s_i$ is exponential, the convergence of (\ref{eq29}) is also exponential.
\item If $s_i$ is bounded and $\lim_{t\rightarrow \infty}|s_i-\Lambda| =O(\varepsilon)$ for some constant $\Lambda$ and small positive constant $\varepsilon$, then the multi-agent system (\ref{eq1}) achieves the following practical output consensus:
\begin{equation}\label{eq4}
\lim_{t\rightarrow \infty}|y_i-y_j|= O(\varepsilon), ~1\leq i\neq j\leq N.
\end{equation}
\end{itemize}}
\noindent \textbf{Proof.} The proof is straightforward from Lemma 1 in Dong (2019) and the linear control theory (Chen, 1998). $\blacksquare$
\vspace{6pt}

Lemma 1 indicates that the consensus of $s_i$, $1\leq i\leq N$, guarantees the output consensus of the multi-agent system (\ref{eq1}). Thus, in the sequel, the protocol $u_i$ will be designed to handle the consensus of $s_i$. Specifically, the information of $s_i$ will be transmitted between each pair of agents.
The dynamics of $s_i$ can be written as
\begin{equation}\label{eq3}
\dot{s}_i=k_1\varrho_{i2}+k_2\varrho_{i3}+\cdots+k_{r-1}\varrho_{ir}+F_{i}(\varrho_i,z_i,\omega_i)+u_i.
\end{equation}
Since the communication network is a digital network, $s_i$ can only be exchanged in discrete-time. Let $T$ be the sampling period and $k=0, 1, \ldots$ the sampling index. Data to be transmitted at time $t=kT$ is then denoted by $s_i(kT)$. What is more, we mention that the sampling period $T$ is also a design parameter in this paper. We have the following lemma.
\vspace{6pt}
\\
\emph{\textbf{Lemma 2.} (Li et al. (2011)) If Assumption A3 holds and $T\in(0, 2/\lambda_N(\mathcal{L}))$, then $\rho_h<1$, where
\begin{equation*}
  \rho_h=\max_{2\leq i\leq N}\left|1-T\lambda_i(\mathcal{L})\right|.
\end{equation*}
Furthermore, if  $T\in (0, 2/(\lambda_2(\mathcal{L})+\lambda_N(\mathcal{L})))$, then $\rho_h=1-T\lambda_2(\mathcal{L})$.}
\vspace{6pt}

For agent $j$, the encoder $\Upsilon_j$ to encode the signal $s_j(t)$ is given by
\begin{equation}\label{eq8}
 \left\{
  \begin{aligned}
   \xi_j(0) &= 0, \\
  \xi_j((k+1)T)&= \beta(kT)\Delta_j((k+1)T)+\xi_j(kT), \\
  \Delta_j((k+1)T)&= q\left(\frac{s_j((k+1)T)-\xi_j(kT)}{\beta(kT)}\right), \\
  & \quad  k=0, 1, \ldots
  \end{aligned} \right.
\end{equation}
where $\xi_j$ is the internal state of $\Upsilon_j$, $\Delta_j$ is the output of $\Upsilon_j$, and $\beta(\cdot)$ is a scaling function to be specified latter. The agent $j$ broadcasts $\Delta_j$ to its neighbors through the digital network. Let agent $i$ be a neighbor of agent $j$. Then agent $i$ receives $\Delta_j$ and uses the following decoder $\Psi_{ji}$ to decode $\Delta_j$ (i.e., to reconstruct $s_j$):
\begin{equation}\label{eq9}
 \left\{
  \begin{aligned}
  \widehat{s}_{ji}(0)&=  0, \\
  \widehat{s}_{ji}((k+1)T)&= \beta(kT)\Delta_j((k+1)T)+\widehat{s}_{ji}(kT), \\
  & \quad k=0, 1, \ldots
  \end{aligned} \right.
\end{equation}
where $\widehat{s}_{ji}$ is the output of $\Psi_{ji}$.

Then for every agent, based on the full-information assumption and the outputs of its own encoder and decoders, the control protocol is given by
\begin{align}\label{eq5}
  u_i(t)=&-\theta_i(t)+\sum_{j\in\mathcal{N}_i}a_{ij}\left(\widehat{s}_{ji}(kT)-\xi_i(kT)\right), \nonumber \\
  & t\in[kT, (k+1)T), ~k=0, 1, \ldots,
\end{align}
where $\theta_i(t)=k_1\varrho_{i2}+k_2\varrho_{i3}+\cdots+k_{r-1}\varrho_{ir}+F_i(\varrho_i,z_i,\omega_i)$.

In the sequel, we will investigate the convergence of the closed-loop system under the full-information-based protocol. Define a compact set $\mathcal{X}\subseteq \mathbb{R}^r$ and let $C_s\geq\max_{\varrho_i\in\mathcal{X}, 1\leq i\leq N}|s_i|$. Recall that the state transformation from $x_i$ to $\varrho_i$ doesn't change the BIBS property of the zero dynamics $\dot{z}_i  = Z_i(\varrho_i,z_i,\omega_i)$. Let $z_i(0)\in \mathcal{Z}_i$ for some compact set $\mathcal{Z}_i\subseteq \mathbb{R}^{n_i-r}$. By Assumption A2, there exists a positive constant $c_{zi}$ such that $\sup_{t\in[0, \infty)}\|z_i\|\leq c_{zi}$ for all $\varrho_i\in \mathcal{X}$. Note that any compact subset of $\mathbb{R}^r\times \mathbb{R}^{n_i-r}$ can be put in the interior of $\mathcal{X}\times \mathcal{Z}_i$.
\vspace{6pt}
\\
\textbf{Theorem 1.} \emph{Consider the closed-loop system composed of the multi-agent system (\ref{eq1}), the encoder (\ref{eq8}), the decoder (\ref{eq9}), and the protocol (\ref{eq5}). Suppose Assumptions A1 to A3 are satisfied, and the initial conditions of the agents $(\varrho_i(0),z_i(0))\in \mathcal{X}\times \mathcal{Z}_i$, $1\leq i\leq N$. Let
\begin{align}
\label{eq15}   T  \in &\left(0, 2/\lambda_N(\mathcal{L})\right),  \\
\label{eq69}   \gamma \in & (\rho_h, 1), \\
\label{eq17}    K \geq  &\left\lfloor K_1(T,\gamma)-\frac{1}{2}   \right\rfloor+1, \\
\label{eq6} \beta(kT)= & \beta_0\gamma^k,
\end{align}
where
\begin{align}
\label{eq79} & K_1(T, \gamma)= \frac{\sqrt{N}T^2\lambda^2_N(\mathcal{L})}{2\gamma(\gamma-\rho_h)}
+\frac{1+2Td^*}{2\gamma},  \\
\label{eq80} & \beta_0> \max\left\{\frac{2T\lambda_N(\mathcal{L})C_s}{\gamma(K+\frac{1}{2})}, \frac{C_{s}}{K+\frac{1}{2}}, \right. \nonumber \\
& \qquad \qquad \quad \left. \frac{2C_s(\gamma-\rho_h)(2\gamma+T\lambda_N(\mathcal{L}))}{T\lambda_N(\mathcal{L})}    \right\}.
 \end{align}
Then the multi-agent system (\ref{eq1}) achieves the output consensus (\ref{eq29}) exponentially.}
\vspace{6pt}
\\
\textbf{Proof.} Under the full-information-based protocol (\ref{eq5}), one has
\begin{equation*}\label{eq23}
  s_i((k+1)T)=s_i(kT)+T\sum_{j\in\mathcal{N}_i}a_{ij}\left(\widehat{s}_{ji}(kT)-\xi_i(kT)\right).
\end{equation*}
The system above falls into the same form of the first-order integrator system considered in Li et al. (2011). Then according to Theorem 3.1 in Li et al. (2011) and the first statement in Lemma 1, one can readily conclude that the multi-agent system (\ref{eq1}) achieves the output consensus (\ref{eq29}) exponentially. $\blacksquare$
\vspace{6pt}

Note that the protocol (\ref{eq5}) is an ``ideal" one since it relies on system full-information, which is unavailable in the general case. However, it provides some insights on the ``ideal" consensus performance the protocol could achieve. We will show in the next section that we can recover the ``ideal" consensus performance by using the device ESO. Towards that end, let the ideal trajectories generated by the full-information-based protocol be superscripted by the symbol $\star$. Then, the dynamics of $s_i^{\star}$ can be specified by
\begin{equation}\label{eq62}
  \left\{
  \begin{aligned}
 & s^{\star}_i((k+1)T)= s^{\star}_i(kT)+T\sum_{j\in\mathcal{N}_i}a_{ij}\left(\widehat{s}^{\star}_{ji}(kT)-\widehat{s}^{\star}_{ij}(kT)\right), \\
  & \qquad \qquad \qquad \quad k=0, 1, \ldots, \\
  & s^{\star}_i(t)= s^{\star}_i(kT)+(t-kT)\sum_{j\in\mathcal{N}_i}a_{ij}\left(\widehat{s}^{\star}_{ji}(kT)-\widehat{s}^{\star}_{ij}(kT)\right),\\
  & \qquad \quad ~ kT<t<(k+1)T.
  \end{aligned}  \right. \\
\end{equation}
The ideal trajectory of $y_i^{\star}$ can be regarded as the output of the $r$th-order linear system
\begin{equation}\label{eq84}
 \left\{
  \begin{aligned}
          \dot{\varrho}^{\star}_i=& A^{\star}\varrho^{\star}_i+B^{\star}u_i^{\star}, \\
                y_i^{\star}=& \varrho^{\star}_{i1}, ~i=1, \ldots, N,
        \end{aligned} \right.
\end{equation}
where $A^{\star}=\left[\begin{array}{c}
  \begin{array}{cc}
   \textbf{0} _{r-1} & I_{r-1}
  \end{array}\\
  \left[0, -k_1,\ldots,-k_{r-1}\right]
  \end{array}\right]$, $B^{\star}=B$, with full-state-feedback based input
\begin{align}\label{eq85}
  u_i^{\star}(t)=& \sum_{j\in\mathcal{N}_i}a_{ij}\left(\widehat{s}^{\star}_{ji}(kT)-\xi^{\star}_i(kT)\right), \nonumber \\
  & t\in[kT, (k+1)T), ~k=0, 1, \ldots
\end{align}
The system (\ref{eq84}) is referred to as the linear counterpart of the uncertain nonlinear multi-agent system (\ref{eq1}).

\section{Consensus Under Output Feedback}

The control design in the previous section is generally not realizable. In this section, we consider the case that $\omega_{i}$ and $f_{im}(\cdot)$, $0\leq m\leq r$, are uncertain, i.e., the external disturbance and agent dynamics are unknown to the designer. In addition, each agent only has access to its first state variable. We propose to use ESOs to estimate the unmeasurable states, and the unknown external disturbance and agent dynamics.  The ESO is built for each agent locally, and then ESO-based local encoder, decoder, and protocol are designed to guarantee the output consensus of the multi-agent system (\ref{eq1}) with limited data rate.

\subsection{ESO-Based Protocol Design}

Let $\varrho_{i,r+1}\triangleq F_i(\varrho_i,z_i,\omega_i)$ be the extended state of agent $i$. The ESO for agent $i$ is then designed as
\begin{equation}\label{eq10}
\left\{
\begin{aligned}
\dot{\hat{\varrho}}_{im}&= \hat{\varrho}_{i,m+1}+\frac{l_{im}}{\varepsilon^m}(y_i-\hat{\varrho}_{i1}), 1\leq m\leq r-1,\\
\dot{\hat{\varrho}}_{ir}&= \hat{\varrho}_{i, r+1}+\frac{l_{ir}}{\varepsilon^{r}}(y_i-\hat{\varrho}_{i1})+u_i, \\
\dot{\hat{\varrho}}_{i,r+1}&=\frac{l_{i, r+1}}{\varepsilon^{r+1}}(y_i-\hat{\varrho}_{i1}),
\end{aligned}
\right.
\end{equation}
where $\hat{\varrho}_i=[\hat{\varrho}_{i1}, \ldots,  \hat{\varrho}_{ir}]^{\rm{T}}$ and $\hat{\varrho}_{i,r+1}$ are the estimates of the agent state $\varrho_i$ and the extended state $\varrho_{i,r+1}$, respectively, $\hat{\varrho}_{i1}(0)$ to $\hat{\varrho}_{i,r+1}(0)$ are set as 0, $\varepsilon<1$ is a small positive constant, and $L_i=[l_{i1},l_{i2}, \ldots, l_{i, r+1}]^{\rm{T}}\in\mathbb{R}^{r+1}$ is selected such that the  matrix $E_i=\left[\begin{array}{cc}
  \begin{array}{c}
   -L_{i}
  \end{array} & \begin{array}{c}
   I _{r-1} \\
   \textbf{0}^{\rm{T}}_{r-1}
  \end{array}
  \end{array}\right]$ is Hurwitz. Since $\varepsilon$ is a small positive constant, the observer (\ref{eq10}) exhibits peaking phenomenon during the initial period. To prevent the peaking from destabilizing the closed-loop system, we employ the well-known saturation technique (Khalil, 2017; Ran, Wang, \& Dong, 2017; Ran, Wang, Dong, \& Xie, 2020) to saturate the outputs of the observer. Recall that we aim to recover the consensus performance under the full-information-based protocol. Let $M_m\geq \sup_{1\leq i\leq N}|\varrho^{\star}_{im}|$, $1\leq m\leq r+1$. Then the outputs of the ESO are bounded by
\begin{equation}\label{eq28}
\overline{\varrho}_{im}=M_m\textrm{sat}\left(\frac{\hat{\varrho}_{im}}{M_m}\right), ~1\leq m\leq r+1.
\end{equation}
Let $\overline{\varrho}_i=[\overline{\varrho}_{i1}, \ldots, \overline{\varrho}_{ir}]^{\rm{T}}$, and the information to be transmitted between each pair of agents be
 \begin{equation}\label{eq12}
\overline{s}_i=k_1\overline{\varrho}_{i1}+k_2\overline{\varrho}_{i2}+\cdots+k_{r-1}\overline{\varrho}_{i, r-1}+\overline{\varrho}_{ir}.
\end{equation}

Similar to the previous section, the encoder $\Upsilon_j$ for agent $j$ is designed as
\begin{equation}\label{eq13}
 \left\{
  \begin{aligned}
   \xi_j(0) &= 0, \\
  \xi_j((k+1)T)&= \beta(kT)\Delta_j((k+1)T)+\xi_j(kT), \\
  \Delta_j((k+1)T)&= q\left(\frac{\overline{s}_j((k+1)T)-\xi_j(kT)}{\beta(kT)}\right), \\
  & \quad  k=0, 1, \ldots
  \end{aligned} \right.
\end{equation}
The decoder $\Psi_{ji}$ for agent $i$ is as follows:
\begin{equation}\label{eq16}
 \left\{
  \begin{aligned}
  \widehat{\overline{s}}_{ji}(0)&=  0, \\
  \widehat{\overline{s}}_{ji}((k+1)T)&= \beta(kT)\Delta_j((k+1)T)+\widehat{\overline{s}}_{ji}(kT), \\
  & \quad k=0, 1, \ldots
  \end{aligned} \right.
\end{equation}
The ESO-based protocol is then given by
\begin{equation}\label{eq14}
 \left\{
\begin{aligned}
  u_i(t)= & -\overline{\theta}_i(t)+\sum_{j\in\mathcal{N}_i}a_{ij}\left(\widehat{\overline{s}}_{ji}(kT)-\xi_i(kT)\right), \\
  &  t\in [kT, (k+1)T), ~k=0, 1, \ldots
 \end{aligned}
 \right.
\end{equation}
where $\overline{\theta}_i(t)=k_1\overline{\varrho}_{i2}+k_2\overline{\varrho}_{i3}+\cdots+k_{r-1}\overline{\varrho}_{ir}+\overline{\varrho}_{i,r+1}$.

\vspace{6pt}
\subsection{Convergence Analysis}

For subsequent use, some notations are defined as $S(kT) =[s_1(kT), s_2(kT), \ldots, s_N(kT)]^{\rm{T}}$,
$\overline{S} (kT) =[\overline{s}_1(kT), \overline{s}_2(kT), \ldots, \overline{s}_N(kT)]^{\rm{T}}$,
$\widehat{\overline{S}}(kT)=[\xi_1(kT), \xi_2(kT),$ $\ldots, \xi_N(kT)]^{\rm{T}}$,
$e(kT) = \overline{S} (kT)-\widehat{\overline{S}}(kT)$,
$\delta(kT) = S(kT)-J_NS(kT)$.

By (\ref{eq3}) and (\ref{eq14}), the evolution of $s_i(kT)$ can be given by
\begin{align}\label{eq32}
& s_i((k+1)T) =  s_i(kT)+ T\sum_{j\in\mathcal{N}_i}a_{ij}\left(\widehat{\overline{s}}_{ji}(kT)-\xi_i(kT)\right) \nonumber \\
  & \quad  +\int_{kT}^{(k+1)T}\left(\sum_{m=1}^{r}k_m\left(\varrho_{i, m+1}(t)-\overline{\varrho}_{i,m+1}(t)\right)\right)\textrm{d}t.
\end{align}
By the structures of the encoder $\Upsilon_j$ and decoder $\Psi_{ji}$, one has $\widehat{\overline{s}}_{ji}(kT)=\xi_j(kT)$, $i\in\mathcal{N}_j$, $1\leq j\leq N$. It follows from (\ref{eq32}) that
\begin{align}\label{eq33}
S((k+1)T) = & S(kT)-T\mathcal{L}\widehat{\overline{S}}(kT)+\Pi_1(kT) \nonumber \\
= & (I-T\mathcal{L})\overline{S}(kT)+T\mathcal{L}e(kT)\nonumber \\
    & +S(kT)-\overline{S}(kT) +\Pi_1(kT),
\end{align}
where $\Pi_1(kT)=[\pi_{11}(kT), \ldots, \pi_{1N}(kT)]^{\textrm{T}}$, with
\begin{align*}
\pi_{1i}(kT)=\int_{kT}^{(k+1)T}\left(\sum_{m=1}^{r}k_m\left(\varrho_{i, m+1}(t)-\overline{\varrho}_{i,m+1}(t)\right)\right)\textrm{d}t.
\end{align*}
Bearing in mind that $\mathcal{L}J_N=J_N\mathcal{L}=0$, one has
\begin{align}\label{eq39}
  &  \overline{S}((k+1)T)-\widehat{\overline{S}}(kT)  \nonumber \\
=   &  S((k+1)T)-\widehat{\overline{S}}(kT)+\overline{S}((k+1)T)-S((k+1)T) \nonumber \\
 =& (I+T\mathcal{L})e(kT)-T\mathcal{L}\delta(kT) \nonumber \\
   & +(I+T\mathcal{L})(S(kT) -\overline{S}(kT)) \nonumber \\
 &+\Pi_1(kT)+\overline{S}((k+1)T)-S((k+1)T).
\end{align}
It then follows from  (\ref{eq33}) and (\ref{eq39}) that
\begin{align}
\delta((k+1)T) = & (I-T\mathcal{L})\delta(kT)+T\mathcal{L}e(kT)  \nonumber \\
\label{eq34} &     +(I-J_N)\left[T\mathcal{L}\left(S(kT)-\overline{S}(kT)\right) \right. \nonumber \\
& \left. +\Pi_1(kT)\right],  \\
 e((k+1)T)=  &\left[\overline{S}((k+1)T)-\widehat{\overline{S}}(kT)\right] \nonumber \\
\label{eq35} & -\beta(kT) Q\left(\frac{\overline{S}((k+1)T)-\widehat{\overline{S}}(kT)}{\beta(kT)}\right),
\end{align}
where $Q([\nu_1, \ldots, \nu_N]^{\rm{T}})=[q(\nu_1), \ldots, q(\nu_N)]^{\rm{T}}$. Note that in the consensus error dynamics (\ref{eq34}), the second term,  $T\mathcal{L}e(kT)$, is caused by the quantization process; the third term, $(I-J_N)\left[S(kT)-\overline{S}(kT)+\Pi_1(kT)\right]$, is caused by the ESO estimation process. Denote
\begin{equation}
\label{eq37} \alpha(kT)=\frac{1}{\beta(kT)}\delta(kT), ~\zeta(kT)=\frac{1}{\beta(kT)}e(kT).
\end{equation}

Before presenting the convergence results of the closed-loop system, we first state the following lemma, which shows the boundedness of $\alpha(kT)$.
\vspace{6pt}
\\
\textbf{Lemma 3.} \emph{Consider the closed-loop system composed of the multi-agent system (\ref{eq1}), the ESO (\ref{eq10}), the encoder (\ref{eq13}), the decoder (\ref{eq16}), and the protocol (\ref{eq14}). Suppose Assumptions A1 to A3 are satisfied, and the initial conditions of the agents $(\varrho_i(0),z_i(0))\in \mathcal{X}\times \mathcal{Z}_i$, $1\leq i\leq N$. Let the sampling period $T$, the parameter $\gamma$, and the quantization parameter $K$  be selected according to (\ref{eq15}), (\ref{eq69}), and (\ref{eq17}), respectively; let the scaling function be selected as
\begin{equation}\label{eq18}
  \beta(kT)=\max\{\beta_0\gamma^k, \sqrt{\varepsilon}\},
\end{equation}
where $\beta_0$ is given by (\ref{eq80}). Then there exists $\varepsilon^{\dag}>0$ such that for all $0<\varepsilon<\varepsilon^{\dag}$ and $k=0, 1, \ldots$, $\|\alpha(kT)\|< \overline{\alpha}$, where $\overline{\alpha}=\max\left\{\frac{2\sqrt{N}C_s}{\beta_0}, \frac{T\sqrt{N}\lambda_N(\mathcal{L})}{2\gamma(\gamma-\rho_h)}\right\}$. Furthermore, the quantizer will never be saturated.}
\vspace{6pt}
\\
\textbf{Proof.} Lemma 3 will be proved recursively. First, consider the case $t=0$. By the fact that $\|\nu\|_{\infty}\leq \|\nu\|\leq \sqrt{N}\|\nu\|_{\infty}$ for any $N$ dimensional vector $\nu$, one has
\begin{equation}\label{eq36}
  \|\alpha(0)\|= \frac{\|\delta(0)\|}{\beta_0} \leq  \frac{\sqrt{N}\|\delta(0)\|_{\infty}}{\beta_0}\leq \frac{2\sqrt{N}C_{s}}{\beta_0}<\overline{\alpha}.
\end{equation}
What is more, for $\zeta(0)$, one has
\begin{equation}\label{eq94}
  \|\zeta(0)\|_{\infty}=\frac{\|\overline{S}(0)-\widehat{\overline{S}}(0)\|_{\infty}}{\beta_0} =0.
\end{equation}

In the sequel, we consider the case $t=(\kappa+1)T$, $\kappa\geq 0$, with the assumption that $\|\alpha(k T)\|< \overline{\alpha}$ and the quantizer is not saturated for all $0\leq k \leq \kappa$. It follows from the definitions of $\alpha(kT)$ and $S(kT)$, and the boundedness of the control signal $u_i$ in the time interval $[0, (\kappa+1)T]$ that $\|\varrho_i\|$ is bounded by some $\varepsilon$-independent positive constant in the time interval $[0, (\kappa+1)T]$.

We define the scaled ESO estimation error $\eta_i=[\eta_{i1},\ldots, $ $\eta_{i,r+1}]^{\rm{T}}\in\mathbb{R}^{r+1}$ with $\eta_{im}=\frac{1}{\varepsilon^{r+1-m}}(x_{im}-\hat{x}_{im})$, $1\leq m\leq r+1$. According to (\ref{eq27}) and (\ref{eq10}), the error dynamics can be written as
\begin{equation}\label{eq43}
\left\{
  \begin{aligned}
  \varepsilon \dot{\eta}_{im} & = \eta_{i, m+1}-l_{im}\eta_{i1}, ~1\leq m \leq r,\\
  \varepsilon \dot{\eta}_{i, r+1}& = \varepsilon \dot{F}_{i}(\varrho_i,z_i, \omega_i)-l_{i, r+1}\eta_{i1}.
  \end{aligned}
  \right.
\end{equation}
By Assumptions A1 and A2, and the boundedness of $\|\varrho_i\|$, one can conclude that in the time interval $[0,(\kappa+1)T]$, $|\dot{F}_i(\varrho_i,z_i,\omega_i)|\leq N_{1i}$ for some $\varepsilon$-independent positive constant $N_{1i}$. Recall that the matrix $E_i$ is Hurwitz by design. Let $P_i\in\mathbb{R}^{(r+1)\times (r+1)}$ be the unique positive definite matrix solution to the matrix equation  $P_iE_i+E_i^{\textrm{T}}P_i=-I_{r+1}$, and define the Lyapunov function candidate $V_i:\mathbb{R}^{r+1}\rightarrow \mathbb{R}$ as $V_i(\eta_i)=\eta^{\textrm{T}}_iP_i\eta_i$. Then one has $\sigma_{i1}\|\eta_i\|^2 \leq  V_i(\eta_i)\leq \sigma_{i2}\|\eta_i\|^2$ and $\left|\frac{\partial V_i(\eta_i)}{\partial \eta_{i, r+1}}\right|\leq 2\sigma_{i2}\|\eta_i\|$, where $\sigma_{i1}$ and $\sigma_{i2}$ are the minimal and maximal eigenvalues of the matrix $P_i$, respectively.  Then by (\ref{eq43}), the time derivative of $V_i(\eta_i)$ satisfies
\begin{align}\label{eq55}
\frac{\textrm{d}V_i(\eta_i)}{\textrm{d} t}=& \frac{1}{\varepsilon}\left(\sum_{j=1}^{r}(\eta_{i, j+1}-l_{ij}\eta_{i1})\frac{\partial V_i(\eta_i)}{\partial{\eta_{ij}}}\right. \nonumber \\ &
\left. -l_{i, r+1}\eta_{i1}\frac{\partial V_i(\eta_i)}{\partial{\eta_{i, r+1}}}\right) +\dot{F}_{i}(\varrho_i,z_i,\omega_i)\frac{\partial V_i(\eta_i)}{\partial{\eta_{i, r+1}}} \nonumber \\
\leq & -\frac{1}{\varepsilon}\|\eta_i\|^2+2\sigma_{i2}N_{1i}\|\eta_i\|  \nonumber \\
\leq & -\frac{1}{\sigma_{i2}\varepsilon}V_i(\eta_i)+\frac{2\sigma_{i2}N_{1i}}{\sqrt{\sigma_{i1}}}\sqrt{V_i(\eta_i)}.
\end{align}
It follows that
\begin{equation}\label{eq46}
  \frac{\textrm{d}\sqrt{V_i(\eta_i)}}{\textrm{d}t} \leq -\frac{1}{2\sigma_{i2}\varepsilon}\sqrt{V_i(\eta_i)} +\frac{\sigma_{i2}N_{1i}}{\sqrt{\sigma_{i1}}}.
\end{equation}
Solving the inequality above and using $\sigma_{i1}\|\eta_i\|^2 \leq  V_i(\eta_i)\leq \sigma_{i2}\|\eta_i\|^2$ again, one has
\begin{align}\label{eq53}
  \|\eta_i(t)\|
   \leq &\sqrt{\frac{\sigma_{i2}}{\sigma_{i1}}}\|\eta_i(0)\|e^{-\frac{1}{2\sigma_{i2}\varepsilon}t}+\frac{2\sigma_{i2}^2N_{1i}\varepsilon}{\sigma_{i1}}.
\end{align}
Note that the right hand side of the inequality above converges to 0 as $\varepsilon\rightarrow 0$ uniformly for any $t\in(0,(\kappa+1)T]$. Therefore, there exists $\varepsilon_0>0$ such that for any $t_{0}>0$ and $\varepsilon\in(0, \varepsilon_0)$, $\|\eta_i(t)\|=O(\varepsilon)$, $\forall t\in[t_{0}, (\kappa+1)T]$. What is more, one can select $M_m\geq \sup_{1\leq i\leq N}|\varrho^{\star}_{im}|$, $1\leq m\leq r+1$, such that the saturations for the outputs of the observer will not be invoked in the time interval $[t_{0}, (\kappa+1)T]$, i.e., $\overline{\varrho}_{im}=\hat{\varrho}_{im}$, $1\leq m\leq r+1$.

With the convergence of the observers, and the assumption that $\|\alpha(k T)\|<\overline{\alpha}$ and the quantizer is not saturated for all $0\leq k \leq \kappa$, we are ready to investigate the bound of $\alpha((\kappa+1)T)$. Let positive integer $k^*$ satisfy $\beta_0\gamma^{k^*-1}> \sqrt{\varepsilon}$ and $\beta_0\gamma^{k^*}\leq \sqrt{\varepsilon}$, i.e., $\beta(kT)=\sqrt{\varepsilon}$ for all $k\geq k^*$. We consider three cases.

\emph{Case 1):} $\kappa+1<k^*$. In this case,  by (\ref{eq39})-(\ref{eq37}), one has
\begin{align}
\alpha((\kappa+1)T)= & \gamma^{-1}(I-T\mathcal{L})\alpha(\kappa T)+\gamma^{-1}T\mathcal{L}\zeta(\kappa T) \nonumber \\
\label{eq60} & +\frac{\Pi_2(\kappa T)}{\beta_0\gamma^{\kappa+1}},\\
\label{eq58} \zeta((\kappa+1) T)= &  \gamma^{-1}\left(\Delta(\kappa T)-Q(\Delta ( \kappa T))\right),
\end{align}
where
\begin{align*}
  \Pi_2(\kappa T) = & (I-J_N)\left[T\mathcal{L}\left(S(\kappa T)-\overline{S}(\kappa T)\right)+\Pi_1(\kappa T)\right], \nonumber \\
   \Delta(\kappa T) = &  (I+T\mathcal{L})\zeta(\kappa T)-T\mathcal{L}\alpha(\kappa T)+\frac{\Pi_3(\kappa T)}{\beta_0\gamma^{\kappa}}, \nonumber \\
   \Pi_3(\kappa T)= &  (I+T\mathcal{L})(S(\kappa T)-\overline{S}(\kappa T)) \nonumber \\
&   +\Pi_1(\kappa T)+\overline{S}((\kappa+1)T)-S((\kappa+1)T). \nonumber
\end{align*}
Let $\max_{0\leq k\leq \kappa}\|\Pi_2(k T)\|/(\beta_0\gamma^{k+1})\triangleq N_2(\varepsilon)$ and $\|\Pi_3(\kappa T)\|/(\beta_0\gamma^{\kappa}) $ $\triangleq N_3(\varepsilon)$. Due to the convergence of the observers and $\beta(kT)\geq \sqrt{\varepsilon}$, one has $N_2(\varepsilon)\rightarrow 0$ and $N_3(\varepsilon)\rightarrow 0$, as $\varepsilon\rightarrow 0$.

Let $J$ be a unitary matrix defined by $J=[\textbf{1}/\sqrt{N},$ $\psi_2,\ldots, \psi_N]$, where $\psi_i^{\textrm{T}}\mathcal{L}=\lambda_i(\mathcal{L})\psi_i^{\rm{T}}$, $2\leq i\leq N$. Let $\widetilde{\alpha}(\kappa T)=J^{-1}\alpha(\kappa T)=J^{\rm{T}}\alpha(\kappa T)$ and decompose $\widetilde{\alpha}(\kappa T)=[\widetilde{\alpha}_1(\kappa T),\widetilde{\alpha}_2(\kappa T)]^{\rm{T}}$ with a scalar $\widetilde{\alpha}_1(\kappa T)$. One can easily verify that $\widetilde{\alpha}_1(\kappa T)=0$. Denote $\widetilde{P}_{\gamma,T}=\gamma^{-1}\textrm{diag}\left\{1-T\lambda_2(\mathcal{L}),\ldots,1-T\lambda_N(\mathcal{L})\right\}$ and $\psi=[\psi_2,\ldots, \psi_N]$. It follows from the definition of $\widetilde{\alpha}_2(\kappa T)$ and (\ref{eq60}) that
\begin{align}\label{eq86}
\widetilde{\alpha}_2((\kappa+1)T)=&\widetilde{P}_{\gamma,T}\widetilde{\alpha}_2(\kappa T)+\gamma^{-1}T\psi^{\rm{T}}\mathcal{L}\zeta(\kappa T)\nonumber \\
& +\frac{\psi^{\rm{T}}\Pi_2(\kappa T)}{\beta_0\gamma^{\kappa+1}}.
\end{align}
By (\ref{eq86}), one can establish the evolution equation from $\widetilde{\alpha}_2(0)$ to $\widetilde{\alpha}_2((\kappa+1)T)$. This together with $\alpha(\kappa T)=\psi\widetilde{\alpha}_2(\kappa T)$ and $\widetilde{\alpha}_2(\kappa T)=\psi^{\rm{T}}\alpha(\kappa T)$ leads to
\begin{align}\label{eq88}
  \alpha((\kappa+1)T)= & \psi[\widetilde{P}_{\gamma,T}]^{\kappa+1}\psi^{\rm{T}}\alpha(0) \nonumber \\
   & +\gamma^{-1}T\psi[\widetilde{P}_{\gamma,T}]^{\kappa}\psi^{\rm{T}}\mathcal{L}\zeta(0)\nonumber\\
   & +\gamma^{-1}T\psi\sum_{i=0}^{\kappa-1}[\widetilde{P}_{\gamma,T}]^i\psi^{\rm{T}}\mathcal{L}\zeta((\kappa-i)T)\nonumber  \\\
   & +\sum_{i=0}^{\kappa}\frac{\psi[\widetilde{P}_{\gamma,T}]^i\psi^{\rm{T}}\Pi_2((\kappa-i)T)}{\beta_0\gamma^{\kappa+1-i}}.
\end{align}
Note that $\|\psi\|=1$, $\gamma\in(\rho_h,1)$, $\|\mathcal{L}\|=\lambda_N(\mathcal{L})$, $\|\widetilde{P}_{\gamma,T}\|\leq \rho_h/\gamma$, and $\|\zeta(0)\|_{\infty}=0< {C_s}/{\beta_0}$ and $\sup_{1\leq t\leq \kappa T}\|\zeta(t)\|_{\infty}\leq 1/(2\gamma)$ since the quantizer is unsaturated before $t=(\kappa+1)T$. By some manipulations, the Euclidean norm of $\alpha((\kappa+1)T)$  is upper bounded by
\begin{align}\label{eq90}
& \frac{2\sqrt{N}C_{s}}{\beta_0}\left(\frac{\rho_h}{\gamma}\right)^{\kappa}+\frac{\sqrt{N}C_sT\lambda_N(\mathcal{L})}{\beta_0\gamma}\left(\frac{\rho_h}{\gamma}\right)^{\kappa} \nonumber \\
&   +\frac{\sqrt{N}T\lambda_N(\mathcal{L})}{2\gamma (\gamma-\rho_h)}\left(1-\left(\frac{\rho_h}{\gamma}\right)^{\kappa}\right)+\frac{\gamma N_2(\varepsilon)}{\beta_0(\gamma-\rho_h)}.
\end{align}
Considering that $\beta_0> \frac{2C_s(\gamma-\rho_h)(2\gamma+T\lambda_N(\mathcal{L}))}{T\lambda_N(\mathcal{L})}$, one has
\begin{align}\label{eq93}
  \|\alpha(\kappa+1)T\| < \frac{\sqrt{N}T\lambda_N(\mathcal{L})}{2\gamma (\gamma-\rho_h)}+\frac{\gamma N_2(\varepsilon)}{\beta_0(\gamma-\rho_h)}.
\end{align}
Hence for sufficiently small $\varepsilon$, $\|\alpha((\kappa+1)T)\|<\overline{\alpha}$ holds.  On the other hand,
\begin{align}\label{eq61}
& \|\Delta(\kappa T)\|_{\infty}  \nonumber \\
\leq   & \|(I+T\mathcal{L})\|_{\infty}\|\zeta(\kappa T)\|_{\infty}+T\|\mathcal{L}\|\|\alpha(\kappa T)\|+\frac{\|\Pi_3(\kappa T)\|}{\beta_0\gamma^{\kappa}} \nonumber \\
< & \frac{1+2Td^*}{2\gamma} +\frac{\sqrt{N}T^2\lambda^2_N(\mathcal{L})}{2\gamma(\gamma-\rho_h)}+N_3(\varepsilon) \nonumber \\
= & K_1(T,\gamma)+N_3(\varepsilon).
\end{align}
Similarly, one can select a sufficiently small $\varepsilon$ such that
\begin{equation}\label{eq91}
  \|\Delta((\kappa+1)T)\|_{\infty} \leq  K+\frac{1}{2}.
\end{equation}
Therefore, the quantizer is unsaturated when $t=(\kappa+1)T$.

\emph{Case 2):} $\kappa+1=k^{*}$. In this case,  one has
\begin{align}
  \alpha((\kappa+1)T)= & \frac{\beta_0\gamma^{\kappa}}{\sqrt{\varepsilon}}(I-T\mathcal{L})\alpha(\kappa T)+\frac{\beta_0\gamma^{\kappa}}{\sqrt{\varepsilon}}T\mathcal{L}\zeta(\kappa T) \nonumber \\
\label{eq95}  & +\frac{\Pi_2(\kappa T)}{\sqrt{\varepsilon}}, \\
\label{eq96}  \zeta((\kappa+1) T)= & \frac{\beta_0\gamma^{\kappa}}{\sqrt{\varepsilon}}\left( \Delta(\kappa T)-Q(\Delta(\kappa T))\right).
\end{align}
Note that $\frac{\beta_0\gamma^{\kappa}}{\sqrt{\varepsilon}}\leq \frac{1}{\gamma}$. Then by conducting a similar analysis as in (\ref{eq86})-(\ref{eq91}), one can conclude that $\|\alpha((\kappa+1)T)\|< \overline{\alpha}$ and the quantizer is unsaturated for $t=(\kappa+1)T$ in this case.

\emph{Case 3):} $\kappa+1>k^{*}$.  In this case, one has
\begin{align}
  \alpha((\kappa+1)T)= & (I-T\mathcal{L})\alpha(\kappa T)+T\mathcal{L}\zeta(\kappa T)+\frac{\Pi_2(\kappa T)}{\sqrt{\varepsilon}}, \\
 \zeta ((\kappa+1)T)= &  \Delta(\kappa T)-Q(\Delta(\kappa T)).
\end{align}
Note that $1<\frac{1}{\gamma}$. Again similar to case 1), one can obtain that $\|\alpha((\kappa+1)T)\|<\overline{\alpha}$ and the quantizer is unsaturated when $t=(\kappa+1)T$.

Therefore, combining Cases 1) to 3) leads to $\|\alpha((\kappa+1)T)\|< \overline{\alpha}$ and the quantizer is unsaturated when $t=(\kappa+1)T$ if $\|\alpha(kT)\|< \overline{\alpha}$ and the quantizer is unsaturated for all $0\leq k\leq \kappa$.
Finally, we arrive at the conclusion that $\|\alpha(kt)\|<\overline{\alpha}$ for all $k\geq 0$ and quantizer will never be saturated. The proof of Lemma 3 is completed. $\blacksquare$
\vspace{6pt}

Now, with Lemma 3, we are in a position to state the convergence results of the closed-loop system.
\vspace{6pt}
\\
\textbf{Theorem 2.} \emph{Consider the closed-loop system composed of the multi-agent system (\ref{eq1}), the ESO (\ref{eq10}), the encoder (\ref{eq13}), the decoder (\ref{eq16}), and the protocol (\ref{eq14}). Suppose Assumptions A1 to A3 are satisfied, and the initial conditions of the agents $(\varrho_i(0),z_i(0))\in \mathcal{X}\times \mathcal{Z}_i$, $1\leq i\leq N$. Let the sampling period $T$, the parameter $\gamma$, the quantization parameter $K$, and the scaling function $\beta(kT)$ be selected according to (\ref{eq15}), (\ref{eq69}), (\ref{eq17}), and (\ref{eq18}), respectively. Then for any $\sigma>0$, there exists $\varepsilon^{\ddag}>0$, which is dependent on $\sigma$, such that for all $0<\varepsilon<\varepsilon^{\ddag}$,
\begin{itemize}
  \item the ESO (\ref{eq10}) recovers the agent state and the unknown dynamics, i.e., there exists $\tau_1(\varepsilon)>0$ satisfying $\lim_{\varepsilon\rightarrow 0}\tau_1(\varepsilon)=0$, such that $\forall t\in[\tau_1(\varepsilon),\infty)$,
     \begin{equation}\label{eq19}
        \left|\varrho_{im}(t)-\hat{\varrho}_{im}(t)\right|\leq \sigma, 1\leq i\leq N, ~1\leq m\leq r+1.
     \end{equation}
  \item the multi-agent system (\ref{eq1}) achieves practical output consensus, i.e.,
    \begin{equation}\label{eq21}
        \lim_{t\rightarrow \infty}\left|y_i(t)-y_j(t)\right|\leq \sigma, ~ 1\leq i\neq j\leq N.
    \end{equation}
  \item the ESO-based protocol (\ref{eq14}) recovers the output consensus performance under the full-information-based protocol, i.e., with same initial conditions,
     \begin{equation}\label{eq22}
        \sup_{t\in[0, \infty)}\left|y_{i}(t)-y^{\star}_{i}(t)\right|\leq \sigma, ~1\leq i\leq N.
     \end{equation}
\end{itemize}}
\noindent\textbf{Proof.} By Lemma 3, for any $\varepsilon\in(0, \varepsilon^{\dag})$, $\|\alpha(kT)\|<\overline{\alpha}$ for all $k\geq 0$. Then following a similar line of the arguments as in (\ref{eq43}) to (\ref{eq53}), one can conclude that there exists $\varepsilon^{\ddag}_1\in(0,\varepsilon^{\dag}]$ such that for all $\varepsilon\in(0,\varepsilon^{\ddag}_1)$, the statement in the first bullet holds.

By the definitions of $\alpha(kT)$ and $\delta(kT)$, and the selection of $\beta(kT)$, for $k\geq k^*$, one has $\|\delta(kT)\|=\beta(kT)\|\alpha(kT)\|<\overline{\alpha}\sqrt{\varepsilon}=O(\sqrt{\varepsilon})$. Thus $|s_i(kT)-s_j(kT)|= O(\sqrt{\varepsilon})$, $1\leq i\neq j\leq N$, for $k\geq k^*$. Furthermore, for any $t\in[kT, (k+1)T)$, one has
\begin{align}\label{eq57}
&  s_i(t)-s_i(kT) =   (t-kT)\sum_{j\in\mathcal{N}_i}a_{ij}\left(\widehat{\overline{s}}_{ji}(kT)-\xi_i(kT)\right) \nonumber \\
  &  \qquad \qquad +\int_{kT}^{t}\sum_{m=1}^{r}k_m\varepsilon^{r+1-m}\eta_{i,m+1}(\nu)\textrm{d}\nu.
\end{align}
Note that the right hand side of the equation above converges to $O(\varepsilon)$. Thus one can conclude that $\lim_{t\rightarrow \infty}|s_i-\Lambda| =O(\varepsilon)$ for some constant $\Lambda$. By Lemma 1, the statement in the second bullet holds.

For $t\geq \tau_1(\varepsilon)$, considering the convergence of the observer, the trajectory of $s_i(t)$ under the ESO-based protocol can be given by
\begin{equation}\label{eq63}
  \left\{
  \begin{aligned}
 &s_i((k+1)T)= s_i(kT)+T\sum_{j\in\mathcal{N}_i}a_{ij}\left(\widehat{s}_{ji}(kT)-\widehat{s}_{ij}(kT)\right) \\
 & \qquad \qquad \qquad +O(\varepsilon), ~k=0, 1, \ldots,\\
 & s_i(t)= s_i(kT)   +(t-kT)\sum_{j\in\mathcal{N}_i}a_{ij}\left(\widehat{s}_{ji}(kT)-\widehat{s}_{ij}(kT)\right) \\
 & \qquad \quad +O(\varepsilon), ~t\in[\tau_1(\varepsilon),\infty)\cap(kT,(k+1)T),
  \end{aligned}  \right.
\end{equation}
Since $\dot{s}_i$ and $\dot{s}_i^{\star}$ are bounded uniformly in $\varepsilon$ and $s_i(0)=s_i^{\star}(0)$, one has
\begin{equation}\label{eq64}
s_i(t)-s^{\star}_i(t)=O(\tau_1(\varepsilon)), ~\forall t\in [0, \tau_1(\varepsilon)].
\end{equation}
Hence $s_i(\tau_1(\varepsilon))-s^{\star}_i(\tau_1(\varepsilon))=O(\tau_1(\varepsilon))$, this together with (\ref{eq62}) and (\ref{eq63}), the continuous dependence of the solutions of differential equations on initial conditions and parameters (see Theorem 9.1 in Khalil (2002)), and the exponential convergence of the equation $s_i((k+1)T)= s_i(kT)+T\sum_{j\in\mathcal{N}_i}a_{ij}\left(\widehat{s}_{ji}(kT)-\widehat{s}_{ij}(kT)\right)$ (i.e., the exponential convergence of $s_i^{\star}(t)$), leads to
\begin{equation}\label{eq65}
  s_i(t)-s_i^{\star}(t)=O(\varepsilon)+O(\tau_1(\varepsilon)), ~\forall t\geq \tau_1(\varepsilon).
\end{equation}
Combining (\ref{eq64}) and (\ref{eq65}) gives
\begin{equation}\label{eq66}
  s_i(t)-s_i^{\star}(t)=O(\varepsilon)+O(\tau_1(\varepsilon)), ~\forall t\geq 0.
\end{equation}
Since $y_i$ is uniquely decided by $s_i$, (\ref{eq66}) indicates (\ref{eq22}). Therefore the statement in the third bullet holds, and this completes the proof of Theorem 2. $\blacksquare$
\vspace{6pt}
\\
\textbf{Remark 2.} The first bullet in Theorem 2 shows the convergence of the ESO.  For sufficiently small $\varepsilon$, the  ESO estimation error and the transient period $\tau_1(\varepsilon)$ can be both made arbitrarily small. The second bullet shows that the proposed ESO-based protocol enables practical output consensus for the multi-agent system (\ref{eq1}). Also note that similar to Li et al. (2011), the  convergence rate of the practical consensus in this paper is also decided by the scaling function $\beta(kT)$. Since $\|\delta(kT)\|<\beta(kT)\overline{\alpha}$ and $\beta(kT)=\max\{\beta_0\gamma^k, \sqrt{\varepsilon}\}$, the practical convergence needs at most $k^*$ steps. The third bullet shows the performance recovery property of the proposed protocol as the trajectories of $y_i$, $1\leq i\leq N$, can be made arbitrarily close to  $y_i^{\star}$, which are generated by the full-information-based protocol. Unlike the existing consensus protocol for multi-agent systems which gives importance mainly to the steady-state consensus error, the proposed protocol also shapes the transient performance. Also note that $y_i^{\star}$ can be regarded as the output of the linear system (\ref{eq84}) with a full-state-feedback control protocol $u_i^{\star}$. Thus the proposed ESO-based protocol is capable of driving the trajectories of the outputs of a higher-order multi-agent system with unknown nonlinear dynamics, external disturbances, and partially measurable states, arbitrarily close to the outputs of its linear counterpart with fully measurable states.
\vspace{6pt}
\\
\textbf{Remark 3.} In this section, since the external disturbance $\omega_i$ and functions $f_{im}(\cdot)$, $0\leq m\leq r$, are unknown, and each agent only has access to its first state variable $x_{i1}$,  an ESO is employed to estimate these information. Then the quantizer uses the ESO estimated information, and consequently the ESO estimation error will be injected into the quantization process (see (\ref{eq39})-(\ref{eq35})). In this case, the scaling function cannot decrease to 0. If $\beta(kT)\rightarrow 0$ as in (\ref{eq6}), the ESO estimation error will be enlarged to infinity during the quantization process, which makes the quantizer saturated. To handle this problem, in Theorem 2, we let the scaling function converge to $\sqrt{\varepsilon}$. Recall that the ESO estimation error is of the order of $O(\varepsilon)$, letting $\beta(kT)\rightarrow \sqrt{\varepsilon}$ guarantees that the ESO estimation error will not make the quantizer saturated.
\vspace{6pt}

From Theorem 2, it can be observed that the number of the quantization levels, $2K+1$, increases when the number of the agents $N$ increases. In particular, $\lim_{N\rightarrow \infty}(2K+1)=\infty$. However, for a practical digital network, the number of the quantization levels is limited. Our next theorem shows that no matter how many agents there are,  the proposed ESO-based protocol guarantees consensus with a fixed number of quantization levels. The lowest number of the quantization levels we achieve is 3 (i.e., $K=1$, a one-bit quantizer).
\vspace{6pt}
\\
\textbf{Theorem 3.} \emph{Consider the closed-loop system composed of the multi-agent system (\ref{eq1}), the ESO (\ref{eq10}), the encoder (\ref{eq13}), the decoder (\ref{eq16}), and the protocol (\ref{eq14}). Suppose Assumptions A1 to A3 are satisfied, and the initial conditions of the agents $(\varrho_i(0),z_i(0))\in \mathcal{X}\times \mathcal{Z}_i$, $1\leq i\leq N$. For any given $K\geq 1$, let the scaling function $\beta(kT)$ be selected according to (\ref{eq18}), and
\begin{align}
\label{eq67}   T  \in &(0, \min\{2/(\lambda_2(\mathcal{L})+\lambda_N(\mathcal{L})), T_m\}),  \\
\label{eq68}   \gamma= & 1-(1-\epsilon_0)T\lambda_2(\mathcal{L}),
\end{align}
where $\epsilon_0\in(0,1)$ and
\begin{align*}
\label{eq83} T_m =& 2K\epsilon_0\lambda_2(\mathcal{L})\left[\sqrt{N}\lambda^2_N(\mathcal{L})+2\epsilon_0\lambda_2(\mathcal{L})d^* \right. \nonumber \\
& +\left. (2K+1)(1-\epsilon_0)\epsilon_0\lambda_2^2(\mathcal{L})\right]^{-1}. \end{align*}
Then the three statements in Theorem 2 hold.}
\vspace{6pt}
\\
\textbf{Proof.} Note that (\ref{eq67}) and (\ref{eq68}) are sufficient conditions for (\ref{eq15}) and (\ref{eq69}), respectively. What is more, it can be verified from the definition of $T_m$ that
\begin{equation}\label{eq70}
  \frac{1}{2} < K_1(T, \gamma)<K+\frac{1}{2},
\end{equation}
which indicates that the condition (\ref{eq17}) is also satisfied.  Therefore, following a similar line of the proof of Theorem 2, one can conclude that the statements in Theorem 3 hold. $\blacksquare$
\vspace{6pt}
\\
\textbf{Remark 4.} Theorem 3 shows that for a connected undirected network of $r$th-order uncertain nonlinear agents with partially measurable states, no matter how many agents there are, one can always design an ESO-based protocol to guarantee consensus with only one bit information exchange between each pair of adjacent agents at each time step. From a theoretical perspective, the proposed approach achieves the lowest data rate via output feedback. For comparison, in the literature, first-order integrator multi-agent systems require at least one-bit data rate (Li et al., 2011);  second-order integrator multi-agent systems require at least two-bit data rate (Li \& Xie, 2012); $r$th-order integrator multi-agent systems require at least $r$-bit data rate (Qiu et al., 2017a); and  $2r$th-order multi-agent systems with oscillator dynamics require at least $r$ to $2r$-bit data rate (Qiu et al., 2017b).
\vspace{6pt}

Finally, we provide some explanations on the parameters selection of the proposed ESO-based protocol to end this section.
\vspace{6pt}
\\
\textbf{Remark 5.} The saturation bounds $M_m$ satisfying $M_m\geq \sup_{1\leq i\leq N}|\varrho^{\star}_{im}|$, $1\leq m\leq r+1$, are selected such that the saturations will not be invoked in the steady state of the observer. The values of $M_1$ to $M_r$ can be obtained by simulating the linear system (\ref{eq84}) with initial condition $\varrho^{\star}_i(0)\in\mathcal{X}$ and the control protocol $u_i^{\star}(t)$. The calculation of $M_{r+1}$ might end up with a conservative value since $F_i(\cdot)$ is unknown. The parameter $C_s$ satisfying $C_s\geq \max_{\varrho_i\in\mathcal{X}, 1\leq i\leq N}|s_i|$ is chosen to prevent the saturation of the quantizer in the initial stage (Li et al., 2011).
\vspace{6pt}
\\
\textbf{Remark 6.} The parameter $\varepsilon$ is important for the successful implementation of the proposed ESO-based protocol. From (\ref{eq53}) and the fact that $\|\eta_i(0)\|\propto \frac{1}{\varepsilon^r}$, one has
\begin{align}\label{eq11}
 \|\eta_i(t)\|\leq & \sqrt{\frac{\sigma_{i2}}{\sigma_{i1}}}\|\eta_i(0)\|\varepsilon^{r+1}+\frac{2\sigma_{i2}^2N_{1i}\varepsilon}{\sigma_{i1}}=O(\varepsilon),
\end{align}
for all $t\geq t_{\varepsilon}$, where $t_{\varepsilon}\triangleq -2\sigma_{i2}(r+1)\varepsilon \ln \varepsilon$. Since $N_{1i}$ is unknown due to the unknown external disturbance and agent dynamics, it is difficult to accurately establish the bound of the ESO estimation error. Therefore, inherent from the previous ESO results (Khalil, 2017), the calculation of the  upper bound of $\varepsilon$ (i.e., $\varepsilon^{\ddag}$) is not straightforward. What is more, for practical control systems, the lower bound of $\varepsilon$ is also limited by some implementation issues, such as noises.  In practice, the value of $\varepsilon$ can be selected by a simple trial and error procedure.  Our experiences and  many ESO application examples in the literature (see Chen, Yang, Guo, and Li (2016) and  Sariyildiz, Oboe, and  Ohnishi (2020), and references therein) indicate that it is generally not difficult to select a satisfactory value of $\varepsilon$.

\section{Example}

\begin{figure}[!t]
  \centering
  \includegraphics[width=0.22\textwidth,bb=15 15 160 100, clip]{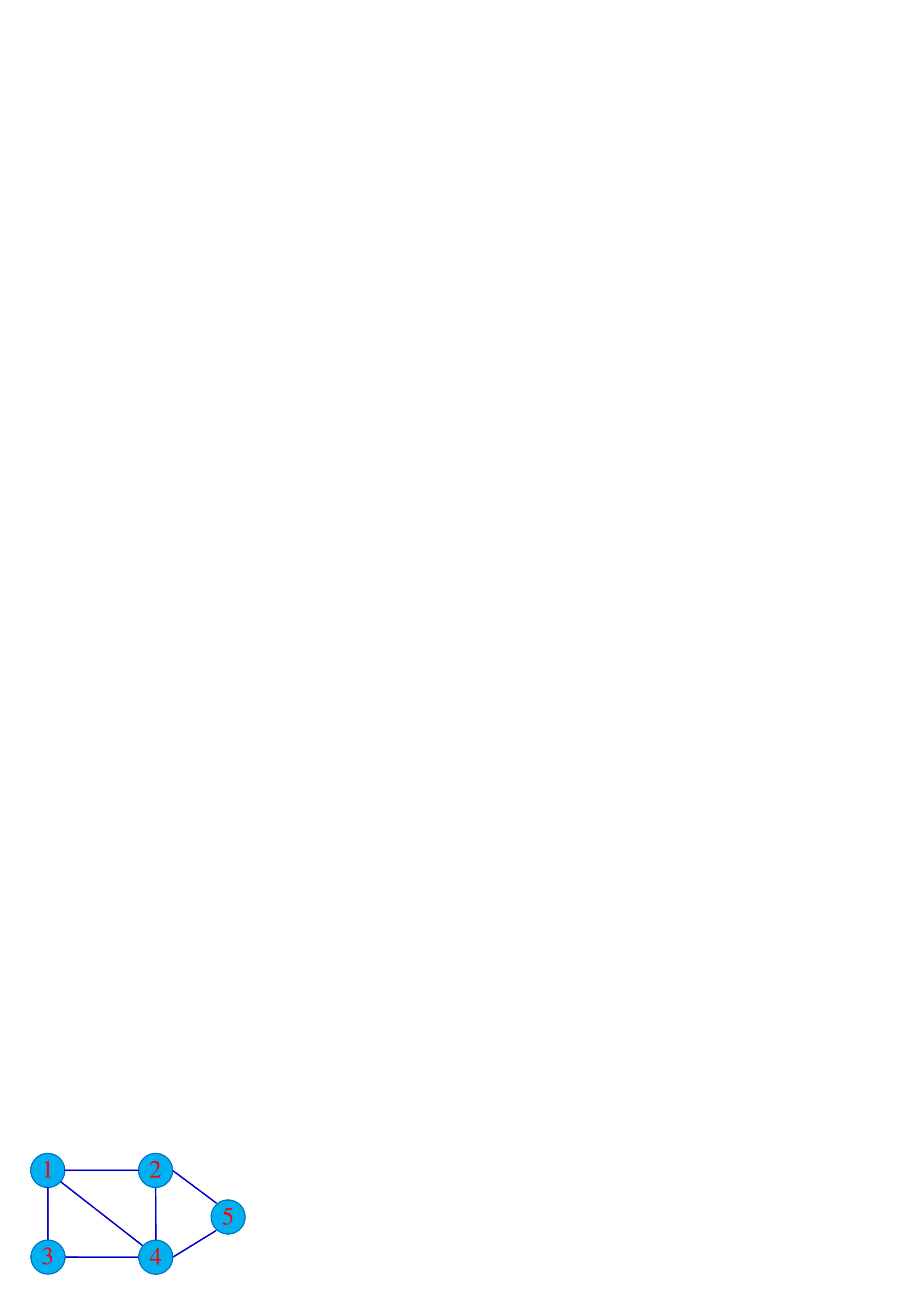}
  \caption{The digital communication network.}\label{fig1}
\end{figure}

In this section, an application example with five pendulum systems is investigated to illustrate the effectiveness of the proposed design scheme. Each pendulum with the inclusion of motor dynamics is described by (Kwan, 1995)
\begin{equation}\label{eq74}
 \left\{
  \begin{aligned}
         \dot{x}_{i1}=&x_{i2}, \\
         \dot{x}_{i2}=&x_{i3}-p_i\sin(x_{i1})-q_i\cos(x_{i1}),\\
         \dot{x}_{i3}=&-x_{i3}+u_i+\omega_{i}, ~1\leq i\leq 5,
        \end{aligned} \right.
\end{equation}
where $x_{i1}$ and $x_{i2}$ denote the angular position and rate of the $i$th pendulum, respectively; $x_{i3}$ denotes the motor shaft angle; $u_i$ is the input motor torque; $p_i$ and $q_i$  are system parameters; and $\omega_{i}$ is the external disturbance. By defining $\varrho_{i1}=x_{i1}$, $\varrho_{i2}=x_{i2}$, and $\varrho_{i3}=x_{i3}-p_i\sin(x_{i1})-q_i\cos(x_{i1})$, the pendulum systems can be rewritten as
\begin{equation}\label{eq75}
 \left\{
  \begin{aligned}
         \dot{\varrho}_{i1}=& \varrho_{i2}, \\
          \dot{\varrho}_{i2}=& \varrho_{i3}, \\
          \dot{\varrho}_{i3}=& -\varrho_{i3}-p_i\sin(\varrho_{i1})-q_i\cos(\varrho_{i1})-p_i\varrho_{i2}\cos(\varrho_{i1})\nonumber \\
          & +q_i\varrho_{i2}\sin(\varrho_{i1})+\omega_{i}+u_i, ~1\leq i\leq 5.
        \end{aligned} \right.
\end{equation}
To make the pendulum systems heterogeneous, we let $p_i=10+i$ and $q_i=2+0.2i$, $1\leq i\leq 5$. The external disturbances are numerically taken as $\omega_{i}=\sin(2t)$.

\begin{figure}[!t]
        \centering
        \begin{subfigure}[b]{0.22\textwidth}
            \centering
            \includegraphics[width=\textwidth,bb=0 0 265 203, clip]{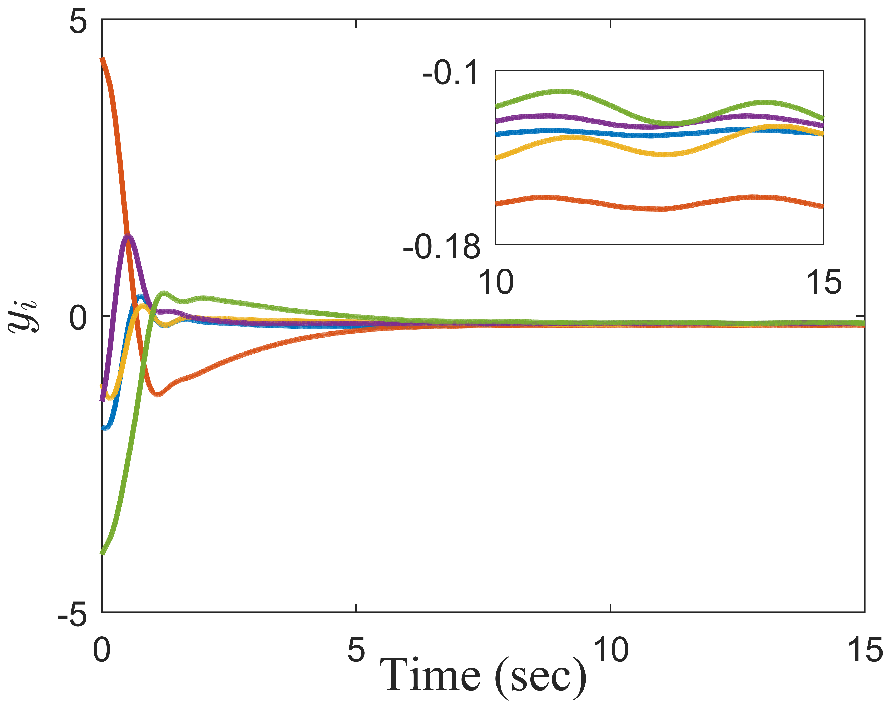}
            \caption[Fig2a]{{$\varepsilon=0.1$.}}
            \label{fig2a}
        \end{subfigure}
        \begin{subfigure}[b]{0.22\textwidth}
            \centering
            \includegraphics[width=\textwidth,bb=0 0 265 203, clip]{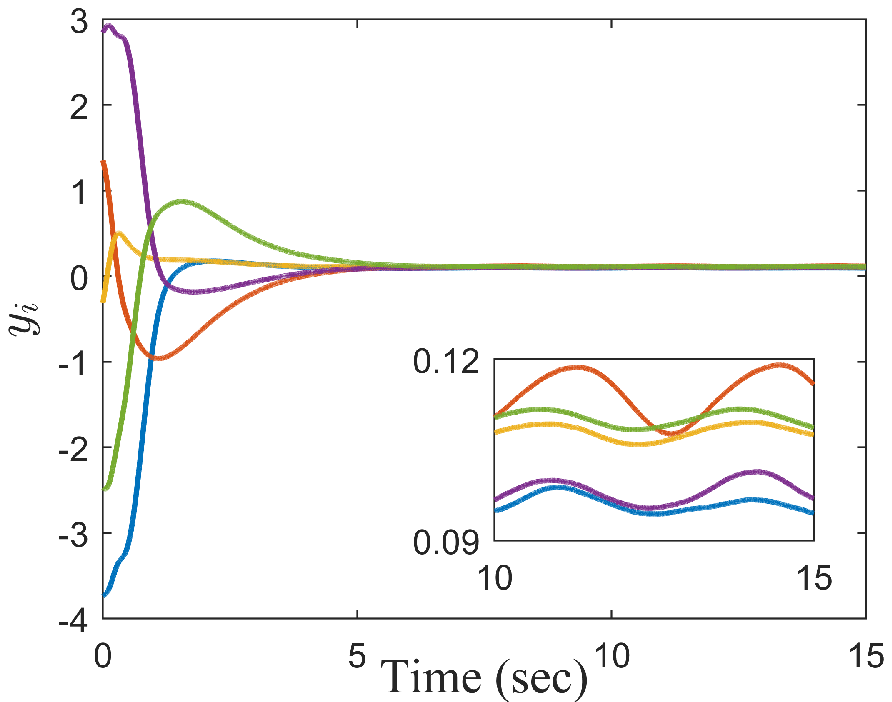}
           \caption[Fig2a]{{$\varepsilon=0.05$.}}
            \label{fig2b}
        \end{subfigure}
        \begin{subfigure}[b]{0.22\textwidth}
            \centering
            \includegraphics[width=\textwidth,bb=0 0 265 210, clip]{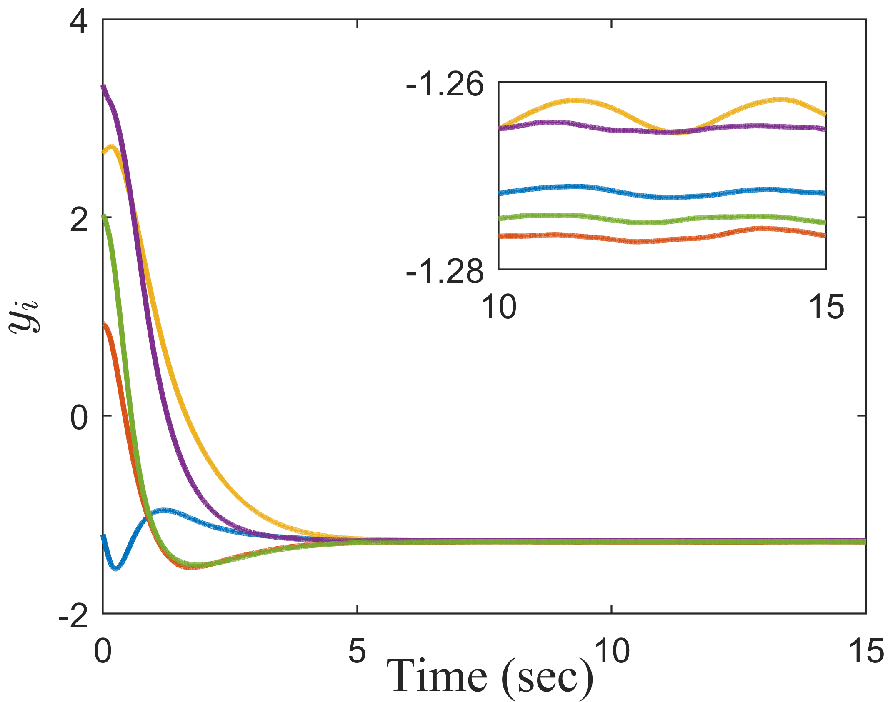}
            \caption[Fig2a]{{$\varepsilon=0.01$.}}
            \label{fig2c}
        \end{subfigure}
                \begin{subfigure}[b]{0.22\textwidth}
            \centering
            \includegraphics[width=\textwidth,bb=0 0 265 210, clip]{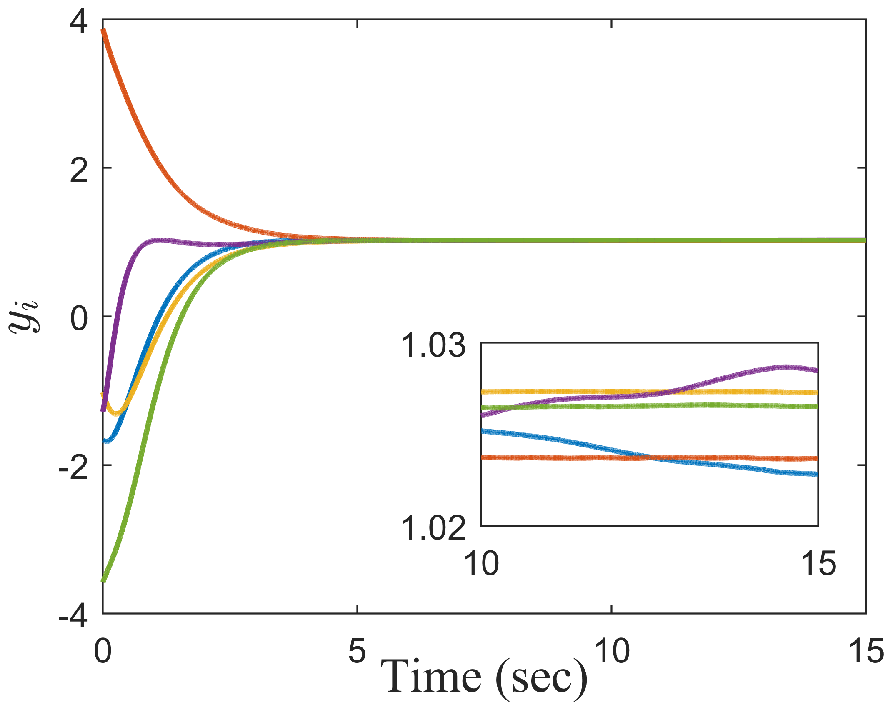}
            \caption[Fig2a]{$\varepsilon=0.001$.}
            \label{fig2d}
        \end{subfigure}
        \caption[]
        {Simulation results of the ESO-based protocol with different values of $\varepsilon$.}
        \label{fig2}
    \end{figure}

\begin{figure}
  \centering
  \includegraphics[width=0.45\textwidth,bb=18 15 435 340, clip]{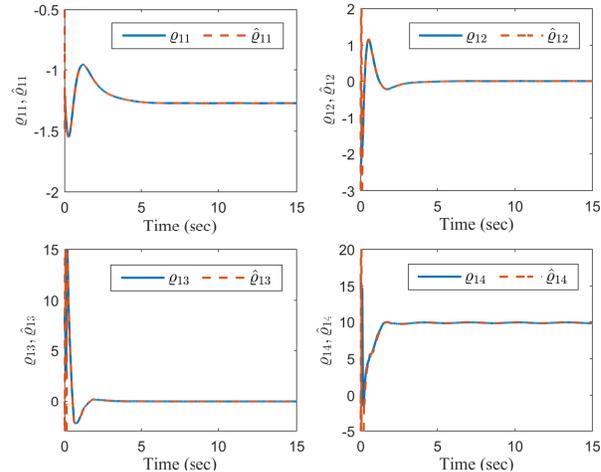}
  \caption{Response of the ESO for agent 1 with $\varepsilon=0.01$.}\label{fig5}
\end{figure}

Fig. \ref{fig1} depicts the communication network among the five pendulums. Let $k_1=k_2=4$. The gains $L_i$, $1\leq i\leq 5$, for the observers are set as $L_i=[4 ~6 ~4 ~1]^{\rm{T}}$, which places the eigenvalues of the matrix $E_i$ at $-1$. The initial conditions of the pendulum systems $x_{im}(0)$, $1\leq i\leq 5$, $1\leq m\leq 3$, are assumed to be randomly distributed in $[-4.5,4.5]$, while the initial conditions of the observers are set as 0. The bounds are set as $M_1=5$, $M_2=5$, $M_3=15$, $M_4=25$, and $C_s=40$.

First, we simulate the higher quantization level case. Select $T=0.05$, $\gamma=0.93$, $K=10$, and $\beta_0=10$, which satisfy all the conditions in Theorem 2. Fig. \ref{fig2}  shows the response of $y_i$, $1\leq i\leq 5$, with different values of $\varepsilon$. One can see that the five agents achieve practical output consensus, and smaller value of $\varepsilon$ leads to smaller  steady-state consensus error. Fig. \ref{fig5} depicts the performance of the ESO for agent 1 with $\varepsilon=0.01$. It can be observed that both the agent state $\varrho_1$ and the unknown nonlinear dynamics $\varrho_{14}\triangleq -\varrho_{13}-p_1\sin(\varrho_{11})-q_1\cos(\varrho_{11})-p_1\varrho_{12}\cos(\varrho_{11})+q_1\varrho_{12}\sin(\varrho_{11})+\omega_{1}$
are well-estimated by the ESO. Also note that the observer exhibits peaking phenomenon during the short transient period.

Then, we investigate the performance of the proposed protocol with one-bit quantizers (i.e., $K=1$). Let $\epsilon_0=0.5$, and by (\ref{eq67}), the sampling period needs to satisfy $0<T<0.0233$. According to Theorem 3, we select $T=0.015$, $\gamma=0.9881$, and $\beta_0=30$. The trajectories of $y_i$, $1\leq i\leq 5$, with $\varepsilon=0.01$, are shown in Fig. \ref{fig7}, which illustrates the practical output consensus of the five agents.

\begin{figure}
  \centering
  \includegraphics[width=0.38\textwidth,bb=15 0 320 240, clip]{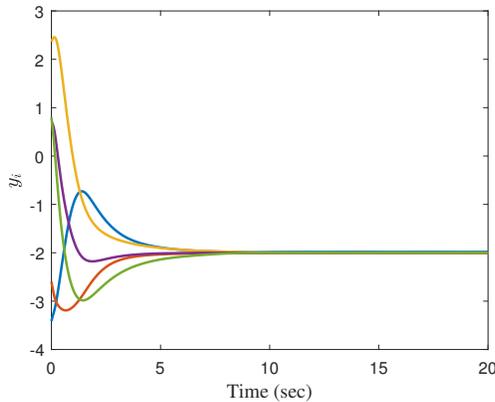}
  \caption{Simulation results of the ESO-based protocol with one-bit quantizers.}\label{fig7}
\end{figure}

\section{Conclusion}

In this paper, we explored the output feedback quantized consensus problem for uncertain nonlinear heterogeneous multi-agent systems. An ESO-based protocol with dynamic encoding and decoding was proposed. The proposed protocol shapes the transient consensus performance, and guarantees a steady-state consensus error that can be made arbitrarily small by tuning the observer parameter. It was also shown that, for a connected undirected network, merely one bit information exchange between each pair of adjacent agents at each time step suffices to guarantee consensus.  The approach developed in this paper provides a practical solution with some intriguing properties to the consensus problem of multi-agent systems with limited data rate. Future research works will be directed at the communication-constrained control problems for multi-agent systems in the presence of network attacks (Feng et al., 2020).

\bibliographystyle{plain}

\begin{thebibliography}{99}


\bibitem{Carli-2010} Carli, R., Bullo, F., \& Zampieri, S. (2010). Quantized average consensus via dynamic coding/decoding schemes. \emph{Internatinoal Journal of Robust and Nonlinear Control}, 20(2), 156-175.

\bibitem{Chen-1998} Chen, C. T. (1998). \emph{Linear System Theory and Design}, 3rd ed. New York, NY, USA: Oxford University Press.

\bibitem{Chen-2016} Chen, W. H., Yang, J., Guo, L., \& Li, S. (2016).  Disturbance-observer-based control and related methods-an overview. \emph{IEEE Transactions on Industrial Electronics}, 63(2), 1083-1095.


\bibitem{Dong-2019}Dong, W. (2019). Consensus of high-order nonlinear continuous-time systems with uncertainty and limited communication data rate. \emph{IEEE Transactions on Automatic Control}, 64(5), 2100-2107.

\bibitem{Feng-2020} Feng, S., Cetinkaya, A., Ishii, H., Tesi, P., \& De Persis, C. (2020). Networked control under DoS attacks: trade-offs between resilience and data rate. \emph{IEEE Transactions on Automatic Control}, published online, DOI: 10.1109/TAC.2020.2981083.

\bibitem{Fie-1973} Fiedler, M. (1973). Algebraic connectivity of graphs. \emph{Czechoslovak Mathematical Journal}, 23(2), 298-305.

\bibitem{Han-2009} Han, J. (2009). From PID to active disturbance rejection control. \emph{IEEE Transactions on Industrial Electronics}, 56(3), 900-906.


\bibitem{Isi-1989} Isidori, A. (1989). \emph{Nonlinear Control Systems}.  New York: Springer-Verlag.


\bibitem{Kash-2007} Kashyap, A., Basar, T., \& Srikant, R. (2007). Quantized consensus. \emph{Automatica}, 43(7), 1192-1203.

\bibitem{Khalil-2002} Khalil, H. K. (2002). \emph{Nonlinear Systems}, 3rd ed. Upper Saddle River, NJ: Prentice Hall.

\bibitem{Khalil-2017} Khalil, H. K. (2017). \emph{High-Gain Observers in Nonlinear Feedback Control}. Society for Industrial and Applied Mathematics (SIAM).

\bibitem{Kwan-1995} Kwan, C. M. (1995). Sliding mode control of linear systems with mismatched uncertainties. \emph{Automatica}, 31(2), 303-307.

\bibitem{Li-2011} Li, T., Fu, M., Xie, L., \& Zhang, J. F. (2011). Distributed consensus with limited communication data rate. \emph{IEEE Transactions on Automatic Control}, 56(2), 279-292.


\bibitem{Li-2012} Li, T. \& Xie, L. (2012). Distributed coordination of multi-agent systems with quantized-observer based encoding-decoding. \emph{IEEE Transactions on Automatic Control}, 57(12), 3023-3037.

\bibitem{Meng-2017} Meng, Y., Li, T., \& Zhang, J. F. (2017). Coordination over multi-agent networks with unmeasurable states and finite-level quantization. \emph{IEEE Transactions on Automatic Control}, 62(9), 4647-4653.

\bibitem{Mu-2015} Mu, X. \& Liu, K. (2015). Containment control of single-integrator network with limited communication data rate. \emph{IEEE Transactions on Automatic Control}, 61(8), 2232-2238.


\bibitem{Qiu-2016} Qiu, Z., Xie, L., \& Hong, Y. (2017a). Quantized leaderless and leader-following consensus of high-order multi-agent systems with limited data rate. \emph{IEEE Transactions on Automatic Control}, 61(9), 2432-2447.

\bibitem{Qiu-2017} Qiu, Z., Xie, L., \& Hong, Y. (2017b). Data rate for distributed consensus of multiagent systems with high-order oscillator dynamics.  \emph{IEEE Transactions on Automatic Control}, 62(11), 6065-6072.

\bibitem{Ran-2017} Ran, M., Wang, Q., \& Dong, C. (2017). Active disturbance rejection control for uncertain nonaffine-in-control nonliner systems.  \emph{IEEE Transactions on Automatic Control}, 62(11), 5830-5836.

\bibitem{Ran-2020} Ran, M., Wang, Q., Dong, C., \& Xie, L. (2020). Active disturbance rejection control for uncertain time-delay nonlinear systems. \emph{Automatica}, 112, 108692.

\bibitem{Ran-2019} Ran, M. \& Xie, L. (2019). Distributed output feedback consensus of uncertain nonlinear multi-agent systems with limited data rate. In \emph{Chinese Control Conference}  (pp. 5733-5739).

\bibitem{Sari-2020} Sariyildiz, E., Oboe, R., \& Ohnishi, K. (2020). Disturbance observer-based robust control and its applications: 35th anniversary overview. \emph{IEEE Transactions on Industrial Electronics}, 67(3), 2042-2053.

\bibitem{You-2011} You, K. \& Xie, L. (2011). Network topology and communication data rate for consensusability of discrete-time multi-agent systems. \emph{IEEE Transactions on Automatic Control}, 56(10), 2262-2275.

\end{thebibliography}

\end{document}